\documentstyle[aps,graphics]{revtex}

\title{Multiresolution analysis for efficient, high precision
all-electron density-functional calculations}

\author{Torkel D.~Engeness$^\dag$ and T.A.~Arias$^\ddag$\\
$\dag$ Department of Physics, Massachusetts Institute of Technology,
Cambridge, MA 02139\\
$\ddag$ Laboratory of Atomic and Solid State Physics, Cornell
University, Ithaca, NY 14853\\
}

\twocolumn

\abstract{ Multiresolution analysis of electronic structure affords
the opportunity to capture the full physics of atomic cores in a
systematically improvable manner.  Applying new techniques, we
demonstrate for the first time that multiresolution analysis of
all-electron calculations within density-functional theory can be
carried out to high precision with a computational effort comparable
to that of the corresponding plane-wave pseudopotential calculation, which
neither captures the full core physics nor is systematically
improvable.  With this approach, we present calculations of
paramagnetic core-level shifts where local density-functional theory
is the sole uncontrolled approximation.  \break \vrule height 15pt
width 0pt Draft date: \today.}

\begin{document}

\maketitle

\section{Introduction}
Over the last several decades, the {\em ab initio} density-functional
approach has proven an accurate, reliable and effective tool for the
study of condensed matter.  It has found application in such diverse
areas as the study of surfaces, point defects, melting, diffusion,
plastic deformation, disorder, catalysis, phase transitions and
chemical reactions\cite{bible}.  In principle, the only approximation
required in the practice of density-functional theory is some model
for exchange and correlation effects.  The fact that the forms for
exchange and correlation are universal and independent of {\em a
priori} knowledge of the physics or chemistry of the systems under
study gives reason for far greater confidence in the predictions of
density-functional calculations than those of their semi-empirical
counter-parts.

Present practice of electronic structure calculation, however, falls
short of this ideal.  All approaches used in current production employ
prior knowledge of chemistry as a criterion for freezing out degrees
of freedom in order to deal with the special demands of representing
physics near the Coulomb singularity of the atomic nucleus.  While
such schemes succeed in making {\em ab initio} studies of complex
phenomena accessible, they make it difficult to improve the
calculations systematically and face the danger of introducing
artificial biases.  For instance, Gaussian basis sets are built
directly from chemical intuition of how atomic orbitals are expected
to polarize\cite{gausschem}.  The linear muffin tin orbital (LMTO)
method \cite{lapwlmto}, the linearized augmented plane wave (LAPW)
method \cite{lapwbook}, and the full potential LAPW (FLAPW) method
\cite{flapw} all use one type of basis set inside of a set of spheres
organized around the nuclei and another type of basis set outside of
the spheres, thus treating physics differently in different regions.
Finally, the pseudopotential, or "frozen core", approximation, ignores
potentially important effects such as core polarization and interferes
with the valence wave functions.  As a result, it is not uncommon for
different groups using different techniques, or even the same
technique, to disagree on the predictions of density-functional
theory.

If one could perform {\em ab initio} calculations with comparable
computational effort, but without the biases of the above approaches
and with the ability to systematically improve the basis, one finally
could access directly the predictions of density-functional theory
without ambiguity.  Also, one could better explore the development of
improved energy functionals, which demands not only better functionals
but also the use of basis sets with highly controlled precision
sufficient to resolve reliably the associated millihartree-level
improvements.

Indeed, wavelet bases do not incorporate any prior knowledge of
electron physics and provide a natural framework for systematic
expansion of the high spatial frequencies which electrons exhibit near
nuclei.  Pioneering {\em all-electron} density-functional calculations
based on multiresolution analysis have been reported, first on
atoms\cite{mgras} and then molecules\cite{rmp,cho}.  Other
applications of multiresolution analysis to related problems include
single-electron problems\cite{prl}, pseudopotential
calculations\cite{chou,tymczak,ivanov} and the solution of Poisson's
equation\cite{goedecker}.  The initial applications to electronic
structure, however, were based on relatively primitive techniques and
were never demonstrated to produce results at the limits of the
accuracy of density-functional theory nor to provide results with an
efficiency comparable to standard techniques such as the
pseudopotential plane-wave approach.  We believe that the reason for
this shortcoming is that standard approaches in the wavelet literature
are designed for applications, such as digital signal processing,
which have far different demands than do continuum problems in the
physical sciences.  In response, we have taken a different tact.
Starting with the fundamental principle underlying wavelet theory,
multiresolution analysis\cite{mallat,meyer,meyer2}, we have developed
a unique set of methods and tools designed specifically for the
treatment of continuum problems\cite{mgras,rmp,ross}.  Below, we
report the results of the use of these new techniques.

\section{Choice of bases} \label{sec:restriction}

\begin{figure}
\begin{center}
\scalebox{0.4}{\includegraphics{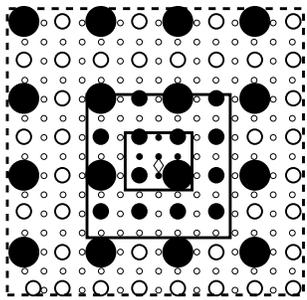}}
\end{center}
\caption{Schematic representation of a two-dimensional restricted
multiresolution analysis of three levels of resolution.
(See text for description.)
}
\label{fig:mra2dmonocube}
\end{figure}

Figure~\ref{fig:mra2dmonocube} illustrates the general structure of
the multiresolution analyses employed in this work.  These begin with
a coarse representation consisting of a uniform orthorhombic grid of
rather extended basis functions (large circles in the figure) placed
in a unit cell with periodic boundary conditions (dashed square).
Additional basis functions of intermediate extent (medium circles) on
a grid of one-half the original spacing then carry details which the
coarse representation does not.  Yet smaller functions (small circles)
on a grid of one-half the spacing of the intermediate grid carry
information for one further scale down.  Repetition of this
construction provides for any desired level of resolution.

With appropriate choice of basis functions, multiresolution analysis
ensures the crucial result which allows the following calculations to
achieve systematic convergence, that such a multiscale basis spans
exactly the same function space as would a uniform basis of functions
on the finest scale.  The basic idea is to insist that each basis
function be expressible {\em exactly} as a linear combination of
translates of itself scaled down by a factor of two,
\begin{equation} \label{eqn:twoscale}
b(\vec x)=\sum_{\vec n} c_{\vec n} b(2 \vec x - \vec n),
\end{equation}
where $b(\vec x)$ is the basis function and $\vec n$ ranges over all
triplets of integers.  The above equation is known as the {\em
two-scale} relation.  Applied recursively, this relation implies that
all coarser functions in the basis are expressible exactly as linear
combinations of functions on the finest scale and, therefore, that the
multiscale basis is equivalent to the basis on the finest scale.  For
a full review, see \cite{rmp}.

The basis consisting of {\em all} functions indicated in the figure
contains precisely as many functions as the equivalent uniform basis
on the finest scale.  Thus, its direct use would represent no savings.
However, the coefficients of the basis functions on the finest levels
of the multiresolution basis represent high frequency details and
hence are significant only in the vicinity of the nucleus (diamond in
the figure).  Consequently, the finest scale functions are necessary
only in the region closest to the nucleus (small solid square), and
the intermediate scale functions are necessary only in a region
extending somewhat further from the nucleus (medium solid square).
{\em Restricting} the multiresolution analysis by maintaining only
those functions (filled circles) within the {\em refinement regions}
for each scale affords significant savings while affecting the final
results negligibly --- shortly below in Section~\ref{subsec:pulay}, we
demonstrate the effects of such restriction to be controllable down to
at least the {\em micro}hartree per atom level.

To describe fully the bases used in our calculations, we must specify
both the original multiresolution analysis and its restriction.  All
multiresolution analyses employed in this work begin with a coarsest
level of one bohr spacing and employ basis functions of the
``third-order {\em semicardinal} type''.  This means that
the functions satisfy the two-scale relation (\ref{eqn:twoscale}) with
the two additional constraints of ``cardinality,''
\begin{equation} \label{eqn:cardinality}
b(\vec n) \equiv \delta_{\vec n,\vec 0}
\end{equation}
(zero value on all integer points but $\vec n = \vec 0$),
and ``exact reconstruction'' of multinomials up to third order,
\begin{equation} \label{eqn:interpolate}
x^\alpha y^\beta z^\gamma \equiv \sum_{\vec n} n_x^\alpha n_y^\beta n_z^\gamma
b(\vec x - \vec n);  \mbox{\ \ $0 \le \alpha, \beta, \gamma, \le 3$}.
\end{equation}
These two additional constraints, respectively, simplify the
computational algorithms considerably and allow third-order
interpolative representation of functions such as the electron
density.  See \cite{rmp} for a full discussion. 

To describe the restrictions employed in this work, we note that they
consist of orthorhombic refinement regions.  Thus,
specification of the restriction reduces to identification of the
dimension ($N_x$,$N_y$,$N_z$) and location of these regions at each
{\em level} of resolution.  Because the refinement regions of a finer
level (higher level number) are always contained within the refinement
region of the next higher level, it is most convenient to identify the
location of a refinement region in terms of the displacement of its
origin from the origin of its {\em parent} region in units of the
parent's spacing ($\Delta_x$, $\Delta_y$, $\Delta_z$).  Using this
approach, for example, Table~\ref{tbl:nmra2dmonocube} specifies the
two-dimensional restriction which Figure~\ref{fig:mra2dmonocube}
illustrates.

In some cases, such as the restriction for the calculation of the
oxygen molecule in Table~\ref{tbl:gridsO2}, a single refinement region
contains multiple {\em children}, regions of the next level of
refinement.  In such cases, we first list the first child with all of
its progeny directly underneath and then proceed to the remaining
children of the original parent.  Ordered and listed with the data as
appear in these tables, such a specification is sufficiently general,
complete and compact that we use it directly as an input data
structure when specifying multiresolution analyses to our software.

\section{Efficacy of multiresolution analysis of electronic structure} \label{sec:acc} 

\subsection{Kohn-Sham equations in a multiresolution basis} 

\def\cG{{\cal G}}
\def\cI{{\cal I}}
\def\cJ{{\cal J}}
\def\cO{{\cal O}}
\def\cL{{\cal L}}
\def\diag{{\mbox{diag}\,}}
\def\Diag{{\mbox{Diag}\,}}
\def\Tr{\,{\mbox{Tr}\,}}

For a full review of the form and solution of the Kohn-Sham equations in a
multiresolution basis, see \cite{rmp}.  Here, we sketch briefly the
key equations which we use to produce the results in this work.
The central quantity in density-functional theory is the energy
expressed in terms of the expansion coefficients $C_{\alpha i}$ of the
Kohn-Sham orbitals,
\begin{equation} \label{eqn:C}
\psi_i(\vec x) = \sum_\alpha C_{\alpha i} b_\alpha(\vec x),
\end{equation}
where $i$ runs over the Kohn-Sham orbitals and $\alpha$ ranges over
the functions in the basis employed in the calculation.  In
\cite{rmp,dftpp}, we give the appropriate expression for this energy
function in an arbitrary basis.  This expression involves standard
matrix operations, such as the trace $\Tr$, and the
basis-dependent matrices
\begin{eqnarray}
\cO_{\alpha \beta} & \equiv & \int b_\alpha(\vec x)^* b_\beta(\vec x)
\, d^3 x \label{eqn:defops} \\ 
L_{\alpha \beta} & \equiv & \int b_\alpha(\vec x)^* \nabla^2 b_\beta(\vec x)
\, d^3 x \nonumber \\
\cI_{p \alpha} & \equiv & b_\alpha(p), \nonumber
\end{eqnarray}
where $\alpha$ and $\beta$ range over all functions in the basis and
$p$ ranges over a set of sample points in real space, which in the
present case are the centers of the wavelet basis functions (filled
circles in Figure~\ref{fig:mra2dmonocube}).  The first two matrices
give the overlap integrals and matrix elements of the Laplacian,
respectively.  The final matrix consists of the values of each basis
function at each sample grid point $p$ so that the product of $\cI$
with a vector of expansion coefficients transforms it to a vector of
values in real space.

In terms of the above matrices, the density-functional expression
for the total energy in terms of the expansion coefficients gathered
into the matrix $C$ is
\begin{eqnarray}
E_{LDA}(C)
& = & \Tr \left( f C^\dagger (-\frac{1}{2} L) C \right) + ({\cI^{-1}}
n)^\dagger \cO v \label{eqn:E} \\
&& + ({\cI^{-1}} n)^\dagger \cO {\cI^{-1}} \epsilon_{xc}(n)  +
\frac{1}{2} \left[ ({\cI^{-1}} n)^\dagger \cO d\right], \nonumber
\end{eqnarray}
Here, $f$ is the occupancy of each orbital (typically 2), $v$ is the
vector of expansion coefficients of the ionic potential,
$$
V_{ion} = \sum_\alpha v_\alpha b_\alpha(\vec x),
$$
$\epsilon_{xc}(n)$ is the usual exchange-correlation energy of
a homogeneous electron of uniform density $n$, 
and $n$ and $d$, respectively, are the electron density evaluated at
the sample points and the expansion coefficients of the Hartree
potential,
\begin{eqnarray}
n & = & \diag\left( \left(\cI C\right) f \left(\cI C\right)^\dagger
\right) \label{eqn:defn} \\
d & = & - 4 \pi L^{-1} \cO \cI^{-1} n, \label{eqn:d}
\end{eqnarray}
where ``$\diag$'' is the operation of forming a vector from the
diagonal elements of a matrix.
The final quantity needed to solve the Kohn-Sham equations is the
derivative of $E_{LDA}$ with respect to the orbital coefficients $C$,
\begin{eqnarray} 
\frac{ \frac{\partial E_{LDA}}{\partial C} }{2f} & \equiv &
-\frac{1}{2} L C +  \left. \cI^\dagger  \Diag \right\{  \label{eqn:grad} \\
&& {\cI^{-\dagger}} \cO {\cI^{-1}}
\epsilon_{xc}(n)  + \left(\Diag \epsilon'_{xc}(n)\right) {\cI^{-\dagger}} \cO
{\cI^{-1}} n \nonumber \\
&& + {\cI^{-\dagger}} \cO v + \left. \left({\cI^{-\dagger}} \cO d\right)
   \right\} \cI C , \nonumber
\end{eqnarray}
where $\Diag a$ represents formation of a diagonal matrix with
diagonal elements given by the components of the vector
$a$.

To aid the reader's interpretation of (\ref{eqn:E}-\ref{eqn:grad}), we
note that these expressions are fully general and applicable not only
to wavelets but also to plane waves.  These expressions, therefore,
represent minor generalizations of the same sequence of operations
found typically in plane-wave density-functional codes.  For instance,
in the computation of the real-space electron density
(\ref{eqn:defn}), the quantity $(\cI C)$ represents the fast Fourier
transformation of the orbitals from coefficient to real space, the
outer-product $\left(\cI C\right) f \left(\cI C \right)^\dagger$ is
then the real-space density matrix, whose diagonal elements ultimately
give the real-space electron density
$n$.\footnote{Eq.~(\ref{eqn:defn}) is a formal expression.
Anticipating the $\diag$ operator, practical software computes only
the diagonal elements and not the full density-matrix.  Similarly, in
Eq.~(\ref{eqn:grad}), rather than forming large matrices such as
$\Diag a$ explicitly, our software computes products such as $(\Diag
a) b$ by multiplying each element of $b$ by the corresponding
element of $a$.}  Similarly, to find the Hartree potential
(\ref{eqn:d}), one first multiplies $\cI^{-1} n$ (inverse Fourier
transforming the real-space density to coefficient space in the plane
wave case), multiplies by $\cO$ (for plane waves, a simple volume
normalization factor) and then by $- 4 \pi$ times the inverse of the
Laplacian matrix ($4 \pi/G^2$ for plane waves).  The terms in
Eqs.~(\ref{eqn:E},\ref{eqn:grad}) each have similar interpretations.

To produce the results reported below, we minimized the expression
(\ref{eqn:E}) over all possible coefficients $C$, using the diagonally
preconditioned conjugate-gradient algorithm augmented with the
analytically continued functional approach\cite{ariaspayne} to handle
the orthonormality constraints on the orbitals.  Combined with
(\ref{eqn:defops}-\ref{eqn:grad}), this describes our calculations fully
and explicitly.

None of the calculations below employ gradient corrected density
functionals.  During the preparation of the manuscript, however, we
were asked to comment on how one would handle the numerical issues
which such functionals raise.  Our general approach to numerical
issues is to note that so long as the evaluation of a given term in
$E_{LDA}$ is exact for any finite expansion of the Kohn-Sham orbitals
(\ref{eqn:C}), then all numerical issues reduce to the quality of the
expansion for the wave functions, which we show below to be
well-controlled in multiresolution bases. 

Therefore, to avoid numerical instabilities associated with finite
differencing to evaluate quantities such as $\nabla n(\vec x)$, we
would evaluate the exact, analytic gradient of the charge density
associated with the orbital expansion (\ref{eqn:C}).  To do
this, we define an additional matrix
$$
\cG_{p \alpha} \equiv \nabla b_\alpha(p),
$$
where $\alpha$ and $p$ range as in the definitions (\ref{eqn:defops}).
The values of $\nabla \psi_i(\vec x)$ at the sample points $p$ are
then
$$
\nabla \psi_i(p) = \left[\cG C\right]_{p i},
$$
so that the gradient of the charge density at these points is {\em
exactly}
\begin{eqnarray*}
\nabla n(p)
& = & f \sum_i \left( \nabla \psi_i(p)^* \psi_i(p) + \psi_i(p)^*
\nabla \psi_i(p) \right) \\
& = & \sum_i \left( \left[\cG C\right]_{pi}^* f \left[\cI C\right]_{pi} + \left[\cI C\right]_{pi}^* f
\left[\cG C\right]_{pi} \right) \\
& = & \left[ (\cI C) f (\cG C)^\dagger + (\cG C)  f (\cI
C)^\dagger \right]_{pp} \\
&&\\
&\Rightarrow &\\
&&\\
\nabla n & = & \diag\left(
(\cI C)^\dagger f (\cG C) + (\cG C)^\dagger f (\cI C) \right).
\end{eqnarray*}
When the exchange-correlation function is evaluated using this
quantity and inverse-transformed with $\cI^{-1}$, the resulting
coefficients, as per the discussion in Section~\ref{sec:compefficacy},
will be exactly the same as would be obtained were the numerical
evaluation carried out in a full multiresolution analysis at arbitrary
resolution without restriction, thereby mitigating any numerical
issues associated with the evalatuation of gradient corrected
functionals on wave functions expanded as in (\ref{eqn:C}). 

\subsection{Systematic convergence} \label{subsec:control}

Present feasible approaches to all-electron calculations complicate
access to the full predictive power of density-functional theory and
the development of new functionals because these methods approach the
atomic cores and the valence regions in fundamentally different ways
and therefore are difficult to bring to convergence in a systematic
manner.  While it has been clear for some time that the ability of
multiresolution analysis to focus resolution in small regions of space
can be exploited to perform all-electron
calculations\cite{mgras,rmp,cho}, the highly systematic nature of
multiresolution analysis and the consequent advantages have yet to be
explored in the context of electronic structure.  We now present the
first demonstration of a multiresolution analysis reproducing
all-electron density-functional results with highly systematic
convergence down to millihartree accuracy.

For this demonstration, we compare results for atoms as calculated
within the local-density approximation as parameterized in\cite{lda}
using both standard techniques and our multiresolution approach.
Atoms provide a convenient test bed because spherical symmetry
produces an effective one-dimensional problem which standard
techniques then readily solve to a precision controllable to better
than 1~nanohartree.  To provide a fair measure of how we expect the
multiresolution method to perform in practice, all calculations with
multiresolution analyses in this work are fully three-dimensional and
without any simplifications.

To underscore that our approach is feasible beyond first and second
row elements, we have chosen calcium for comparison.  To represent the
charge distribution of the calcium nucleus (and all nuclei in this
work), we use the linear combination of three Gaussian distributions
with the widths and charge contents appearing in
Table~\ref{tbl:teter}.  This charge distribution reproduces the atomic
electron eigenvalues to within 0.5~mH\cite{teternuc}, an accuracy
which is systematically improvable through a renormalization
approach\cite{lepage}.  All calculations and comparisons below are
made for this form of nuclear charge distribution.  As described
above, the multiresolution analysis begins with a spacing of 1~bohr on
level zero.  To explore the convergence toward the final solution, we
perform self-consistent calculations with from seven to nine levels
of refinement, for a resolution of down to $1/2^9\sim 0.002$~bohr.
Table~\ref{tbl:Cagridstruct} summarizes the restrictions employed for
the calculation.

\begin{figure}
\begin{center}
\scalebox{0.4}{\includegraphics{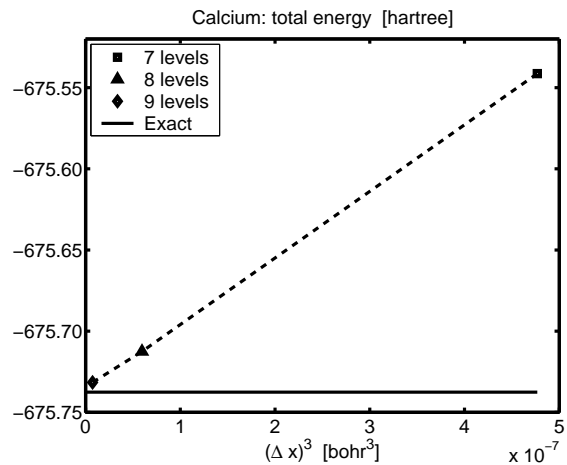}}
\end{center}
\caption{Error in total all-electron energy within the local density
approximation, $\Delta E$, of a multiresolution calculation of an
isolated calcium atom at 7, 8 and 9 levels of resolution (squares)
below a top level of 1~bohr spacing.  The horizontal axis gives the
cube of the resolution on the finest scale $(\Delta x)^3$.}
\label{fig:Ca}
\end{figure}

Figure~\ref{fig:Ca} shows, as a function of the {\em cube} of the
spacing on the finest level, the difference between the total energy
calculated with the multiresolution analysis and the highly precise
solution of the equivalent one-dimensional problem.  When we employ
nine levels of refinement, the absolute error is 6~mH, one part in
$10^5$ of the total energy.

The simple, linear approach of the error toward zero in this
plot is a direct consequence of the fundamental nature of the basis.
The multiresolution analysis which we employ satisfies the two-scale
relation and, thus, is equivalent to a uniform basis on the finest
scale of resolution {\em by construction}.  The linear behavior
evident in the figure, therefore, can be understood {\em a priori}\,
as a direct consequence of the third-order nature of the semicardinal
basis.  The only other method which provides this level of analytic
understanding of the errors as a function of the size of the
calculation is the plane-wave approach.  However, large scale
plane-wave calculations are only feasible in practice with
pseudopotentials and therefore do not systematically converge to the
correct all-electron result.  Multiresolution analysis, therefore, is
unique among {\em feasible} all-electron approaches in allowing for a
high degree of {\em a priori}\, understanding of and systematic control
over errors.

\subsection{Uniformity of description of space} \label{subsec:pulay} 

Present methods demonstrated to be viable for all-electron
calculations all treat space near the atomic nuclei differently from
the interatomic regions.  This inherently biases the description
toward the location of the atoms and complicates the evaluation of
forces on the atoms\cite{pulay_orig}.  Surprisingly, even under
restriction, this effect is nearly absent in multiresolution analysis
because of the unique way in which multiresolution analysis represents
information.

In direct contrast to the expansion coefficients in finite-element
bases or finite-difference calculations, which are in proportion to
the {\em value} of the represented function in the corresponding
region of space, the expansion coefficients of the finer-scale
functions in a multiresolution analysis represent only the
higher-frequency {\em details} in the corresponding region.  The
expansion coefficients of the finer-scale functions which are
restricted from a multiresolution basis (empty circles in
Figure~\ref{fig:mra2dmonocube}), unlike the coefficients which would
appear in those regions in traditional approaches, therefore can be
made to vanish controllably by expanding the corresponding refinement
regions (squares in Figure~\ref{fig:mra2dmonocube}) outward from the
nuclei.  As the coefficients dropped from the expansion vanish, the
restricted multiresolution analysis becomes indistinguishable from the
full analysis,  and the underlying description of space therefore
becomes uniform.

To demonstrate this, we consider the calculation of the total energy
of a single atom (aluminum) at different locations in space.  For
these calculations, we employ a cubic cell of dimension 48 bohr with
seven levels of refinement, each consisting of a cubic array of $48^3$
basis functions.  We begin with the atom centered on a basis function
from the coarsest scale and then move the atom along a coordinate axis
in increments so that the nucleus falls directly on the centers of
basis functions, eventually sampling functions from each of the eight
levels in the calculation.  As the atom moves, we add or remove
functions from the basis to maintain the atom at the center of each
refinement grid while conserving the number of basis functions on each
level.

\begin{figure}
\begin{center}
\scalebox{0.4}{\includegraphics{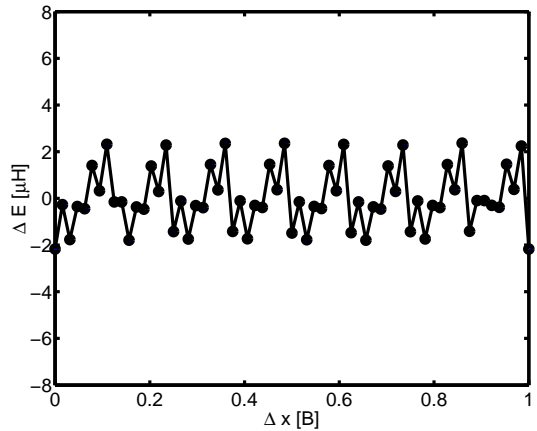}}
\end{center}
\caption{Fluctuation ($\Delta E$) in total all-electron energy of an
isolated aluminum atom as a function of distance ($\Delta x$) of the
nucleus from the center of a basis function on the coarsest scale.  Note
that vertical scale is in {\em micro}hartree.}
\label{fig:nopulay}
\end{figure}

Figure~\ref{fig:nopulay} shows the results of the above calculations.
Despite the switching on and off of the basis functions and the
radically changing nature of the basis function on which the nucleus
falls (ranging by over two orders of magnitude in length scale), the
root-mean-square fluctuation in the energy is less than 3 {\em
micro}hartree, which we could reduce yet further by expanding the
extent of the refinement regions.  In terms of Pulay-force
corrections\cite{pulay_orig}, the maximum fluctuation in energy as we
move the atom by 1/64~bohr (the smallest step for which the basis
actually changes) is 2.0$\mu$H, corresponding to a maximum average
Pulay force of $\sim 128$~$\mu$H/B $\approx$ 0.007~eV/\AA.  This
remarkable result makes the multiresolution approach an extremely
attractive tool in calculations where the atoms move, such as
structural relaxation or molecular dynamics.

\subsection{Description of chemical bonding} \label{subsec:bonding}

To investigate the efficacy of the multiresolution approach in the
description of chemical bonding, we consider the oxygen molecule.
Given the paramagnetic nature of this molecule, we carry out the
calculation within the local spin-density approximation (LSDA) as
parameterized in \cite{lsda}.  To ensure sufficient isolation of the
molecule from its periodic images, we employ a supercell of dimensions
14.8\AA$\times$12.7\AA$\times$12.7\AA.  We then orient the molecule
along the long axis of the cell, place the nuclei at their
experimental separation, and employ the multiresolution analysis in
Table~\ref{tbl:gridsO2}.

For comparison, we also performed, within the
same geometry and parameterization of the local spin-density
functional, a plane-wave calculation within a periodic supercell of
identical dimensions using a pseudopotential optimized for convergence
according to the procedure of Rappe {\em et al.}\cite{rappe}.  This
pseudopotential required a plane wave cut-off of 35~hartree to
converge the total pseudo-energy to within 0.10~hartree.  Given this
level of convergence in the total energy and the fact that the
fractional error in the variational total energy scales like the
square of the fractional errors in non-variational quantities such as
the eigenvalues, we estimate the individual eigen-energies to be
converged to within about 0.03~hartree at this plane-wave
cut off.

\begin{figure}
\begin{center}
\scalebox{0.4}{\includegraphics{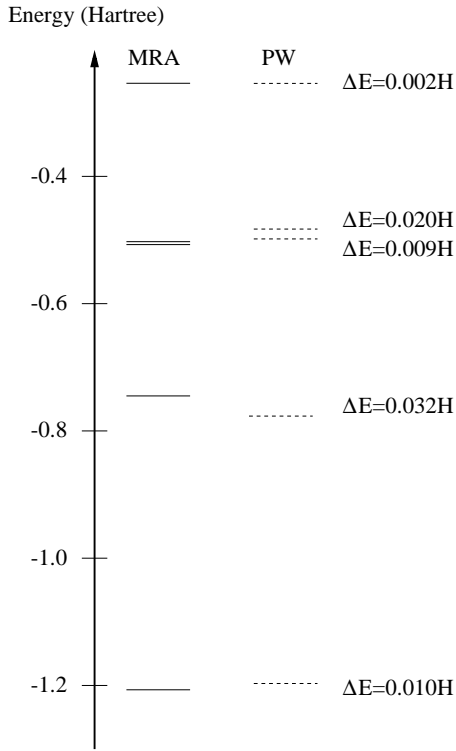}}
\end{center}
\caption{Kohn-Sham eigenvalues for valence states of the oxygen
molecule: results of all-electron multiresolution analysis calculation
(MRA) and plane-wave pseudopotential calculation (PW).}
\label{fig:O2}
\end{figure}

Figure~\ref{fig:O2} compares the Kohn-Sham eigenvalues from the
pseudopotential calculation with those from multiresolution analysis.
Within the expected convergence of the pseudopotential calculation,
the agreement is complete.  These results illustrate that the
multiresolution approach gives a description of bonding at least as
reliable as the standard pseudopotential approach.  Because the issue
of transferability remains an open question in pseudopotential theory,
this result also lends greater credence to the pseudopotential which
we have constructed for these calculations.  We envisage the use of
multiresolution analysis in this way to aide in the construction of
pseudopotentials with greater transferability.

\section{Computational Efficacy} \label{sec:compefficacy}

The computational demands of wavelet-based all-electron calculations
in previous works were so extreme that the literature has yet to
explore the practicality of the computations.  The calculations
presented in this work are the first to apply a series of new
techniques\cite{rmp,ross,torkel} to achieve high performance in
all-electron problems, placing us in a position to account in detail
for the computational effort that our calculations demand and thereby
provide a standard against which the performance of future wavelet
calculations can be compared.  We find that our new techniques make
the demands of all-electron calculations sufficiently comparable to
those of frozen-core pseudopotential calculations of the same
precision to make wavelet calculations a practical alternative.

Rather than using techniques developed for other classes of
application, the techniques which we use here are tailored to the
specific demands of continuum problems in the physical sciences.  Our
basis, for instance, technically is not a wavelet basis because its
dual consists of sets of Dirac-delta functions.  These delta
functions, however, are precisely what are needed to recover {\em
exact} representations from samples of physical fields at extremely
limited numbers of points in real space\cite{ross}.  In the present
context, this means that the operations $\cI^{-1}$ appearing in
(\ref{eqn:E},\ref{eqn:d},\ref{eqn:grad}) always yield precisely the
same results as would be obtained were the calculation carried out
with a full, {\em unrestricted} multiresolution analysis of {\em
arbitrary} resolution.  (See \cite{rmp,ross}.)  As evident in
(\ref{eqn:E},\ref{eqn:d},\ref{eqn:grad}), such operations are key in
the effective treatment of non-linear couplings such as exchange and
correlation effects and evaluation of the Hartree potential.  Without
this profound result, accurate evaluation of such effects requires
evaluation of quantities at a much larger set of points in real space,
thereby slowing the calculation considerably.  To go along with these
new methods for the transforms ($\cI$, $\cI^{-1}$ and related
operations), we have also developed a class of $O(N)$ methods for
multiplication by the operators $\cO$ and $L$ with similar exactness
properties\cite{rmp,ross}.  Finally, to further improve the
performance of the calculations, we have developed special
cache-optimized algorithms for all of these methods\cite{torkel}.

As a case study, we consider in detail the computational demands of
the calculations of the electronic structure of the oxygen molecule
presented in Section~\ref{subsec:bonding} above.  To make the
comparison with a pseudopotential calculation meaningful, we choose a
restriction of the multiresolution analysis (Table~\ref{tbl:gridsO2})
and a plane-wave cut off (35~H) which achieve similar convergence in
the total energy: 0.08~H and 0.10~H, respectively.

\begin{figure}
\begin{center}
\scalebox{0.4}{\includegraphics{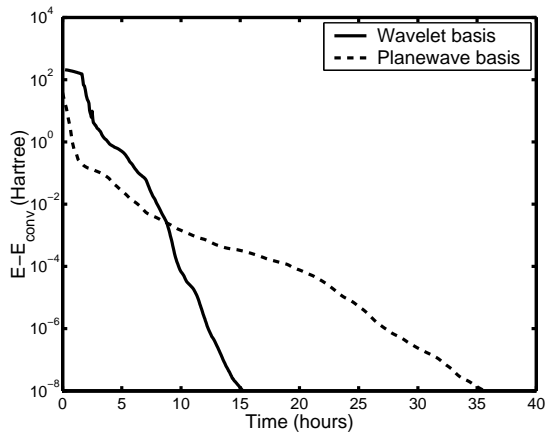}}
\end{center}
\caption{Convergence of total energy of the oxygen molecule as a
function of actual wall-clock time on a 400~MHz Pentium~II PC for
basis sets reaching a similar level of convergence:
plane-wave pseudopotential calculation (dashed curve), all-electron
multiresolution analysis calculation (solid curve).}
\label{fig:o2convergence}
\end{figure}

Figure~\ref{fig:o2convergence} presents the convergence of the total
energy as a function of actual wall-clock time on a 400~MHz Pentium~II
PC both for our wavelet calculation and for a plane-wave
pseudopotential calculation performed with the highly optimized DFT++
code\cite{dftpp}.  Despite the fact that the pseudopotential
calculation freezes out the physics of the core, the absolute
convergence for the pseudopotential calculation is slower than that of
the all-electron wavelet calculation.  These results are the first to
establish multiresolution analysis as a {\em computationally viable},
unbiased and systematic approach to the calculation of all-electron
systems.

Several advances have taken place to make possible this favorable
comparison.  To place these in context, we consider the time required
for the single most time consuming {\em basis dependent} operation in the
two calculations, the Laplacian for the wavelet case and the Fourier
transform for the plane-wave case.  In general, the total time required to
perform such operations is
\begin{equation} \label{eqn:time}
T=N_A \frac{N_B N_{F/B}}{N_{F/T}},
\end{equation}
where $N_A$ is the number of times the corresponding operation is
performed while reaching the desired level of convergence, $N_B$ is
the number of functions in the basis, $N_{F/B}$ is the number of
floating point operations needed to process each basis function during
the operation, and $N_{F/T}$ is the number of floating point
operations which the algorithm for the operation achieves per unit
time.

Table~\ref{tbl:performance} presents these quantities for calculations
which achieved the convergence of 10$^{-8}$~H in
Figure~\ref{fig:o2convergence}.  The table confirms the finding
established in earlier works that the total number of basis functions
$N_B$ required for both calculations is similar\cite{mgras,rmp}.

The new algorithms which we employ\cite{rmp,ross} exploit the local
nature of the multiresolution basis to produce a data flow pattern
which now is sufficiently simple to reduce $N_{F/B}$ from that of our
previous calculations\cite{mgras} by several orders of magnitude.  To
multiply by $\cI$, for instance, we apply the two-scale relation
(\ref{eqn:twoscale}) recursively until we reach an expansion in terms
of functions on the finest scale, which, by virtue of
(\ref{eqn:cardinality}), gives the values of the function at each
point $p$.  Eq.~(\ref{eqn:twoscale}) implies that each step in this
recursion is simply a three-dimensional convolution by the sequence
$c_{\vec n}$.  Reference~\cite{rmp} presents methods for wavelet bases
which similarly decompose multiplication by all remaining basis
dependent matrices ($\cO$, $L$, $\cI^{-1}$, $\cI^\dagger$,
$\cI^{-\dagger}$) into three-dimensional convolutions for wavelet
bases.  For our choice of basis, these convolutions are short and
separable, leading to an operation count with a significant advantage
over that of the fast Fourier transform: $N_{F/B}\approx 250$ versus
$N_{F/B}\approx 75 \log_2 (15 N)$ (including appropriate factors for
mapping the plane-wave sphere into the Fourier transform box).
Table~\ref{tbl:performance} shows that for the present calculations,
the wavelet $N_{F/B}$ is superior by a factor greater than six.

The simpler communication pattern of separable three-dimensional
convolutions also allows us to develop special algorithms with
significantly improved cache performance and thus processing rates
$N_{F/T}$ nearly an order of magnitude improved over that achieved in
previously reported wavelet-based electronic structure calculations.
Table~\ref{tbl:performance} shows that, in fact, we now can achieve
floating point operation rates $N_{F/T}$ somewhat superior to those of
the highly tuned FFTW Fourier transform package\cite{fftw}, which we
employ in the plane wave calculations.  Finally, we note as a possible
area for future research that the only area where the plane wave
calculation holds a significant advantage is in the number of
applications required of the rate limiting operator $N_A$.

The result of all of the above factors is that the plane wave and
wavelet calculations require comparable (within a factor of three)
amounts of time $T$ in their most time consuming basis-dependent
operators.  The remaining components of the calculations do consume
significant amounts of time and so the total run times
(Figure~\ref{fig:o2convergence}) are significantly longer.  In the
wavelet case, our data for the Laplacian operator accounts for about
one-half of the total run time.  The additional operations in the
plane wave calculations consume a much more significant fraction of
the total time.  As a result, in this particular case it turns out
that the final total run time for the pseudopotential calculation,
which does not include the physics of the core explicitly, is in fact
noticeably longer than that for the corresponding wavelet calculation.

\section{Core-level physics}

One great advantage of multiresolution all-electron calculations over
their pseudopotential counterparts is the direct access which they
give to the full physics of the atomic cores in an unbiased
representation.  A promising area of application for such calculations
is their use as an aid to the interpretation of experimental
techniques which measure environmentally dependent shifts in the core
states, such as electron energy loss spectroscopy
(EELS)\cite{cornellfriends} and electron spectroscopy for chemical
analysis (ESCA)\cite{ESCAbook}.  As a simple demonstration of the
physics accessible through this approach, we consider the
environmental dependence of the oxygen $1s$ state in the water and the
diatomic oxygen molecules.

\begin{figure}
\begin{center}
\scalebox{0.4}{\includegraphics{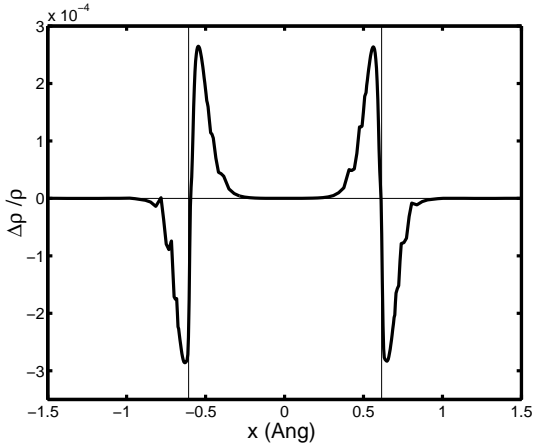}}
\end{center}
\caption{Difference, along the symmetry axis of the molecule, in total
charge density of the oxygen $1s$ states (both spin channels) in going
from atomic oxygen to the oxygen molecule.  Values expressed as
fractions of the peak density of the $1s$ state in atomic oxygen.
Vertical lines indicate locations of the oxygen nuclei.}
\label{fig:O2_O1s}
\end{figure}

For diatomic oxygen, which is paramagnetic, we use the results of the
local spin-density calculations described above in
Section~\ref{subsec:bonding}.  For the water molecule and the
reference atomic states, we employ the local density approximation as
parameterized in~\cite{lda}.  For all calculations, we place the
nuclei at their experimentally known locations and employ the
multiresolution analyses which
Tables~\ref{tbl:gridsO2}~and~\ref{tbl:gridsH2O} summarize.  The
general, simple structure of multiresolution analysis allows this
brief description to specify the calculations completely.

\begin{figure}
\begin{center}
\scalebox{0.4}{\includegraphics{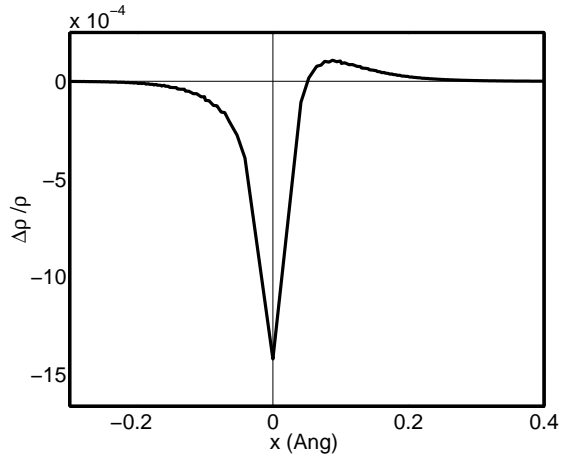}}
\end{center}
\caption{Difference in charge density of the oxygen $1s$ state in
going from atomic oxygen to the water molecule along a line directly
toward one of the protons.  Values expressed as fractions of the peak
density of the $1s$ state in atomic oxygen.  Vertical line indicates
the location of the oxygen nucleus.} \label{fig:H2O_O1s}
\end{figure}

Our results for the deformation of the $1s$ state of oxygen in the
above molecules appear in
Figures~\ref{fig:O2_O1s}~and~\ref{fig:H2O_O1s}.
Figure~\ref{fig:O2_O1s} shows the change in the total electron density
in the oxygen $1s$ states (sum both over spin channels and over
bonding and anti-bonding states) in going from atomic oxygen to the
oxygen molecule.  Despite being quite small (at the 0.03 \% level),
the core polarization effect is quite clearly defined in our
multiresolution analysis, which we emphasize builds no {\em a priori}\,
knowledge of this polarization into the calculation.  Interestingly,
we find that, despite the buildup of repulsive valence charge in the
bond between the oxygen atoms, the decreased screening of the nuclei
is sufficient to produce local electric fields which polarize the core
electrons {\em toward} the center of the molecule.

Figure~\ref{fig:H2O_O1s} shows the change in the electron density of
the oxygen $1s$ state in going from the oxygen atom to the water
molecule.  Again, the deformation of the $1s$ states is quite clearly
described in our multiresolution analysis despite being relatively
small (0.16\%, somewhat larger than in the oxygen molecule).  In the
water molecule, the electrons associated with the hydrogen atoms are
stripped away and distributed in an approximately spherical shell of
charge around the oxygen atom.  Similarly to the water molecule, our
calculation finds that this lessens the shielding of the protons and
leads to the net attraction of the oxygen $1s$ state toward each of
the protons which is evident in the figure.  The figure also shows
that, in the water molecule, the displacement of the $1s$ state leads
to a net decrease of its amplitude on the nucleus.  To quantify the
impact of these core polarization effects on matrix elements
describing transitions from the core to higher states, we have
calculated the oxygen $1s$ contribution to the dipole moment of the
water molecule, finding 0.0003~electron-bohr.  As a point of
reference, the net dipole moment which we compute for the entire
molecule is 0.731~electron-bohr, in excellent agreement with the
tabulated experimental value of 0.729~electron-bohr\cite{crc}.

\begin{figure}
\begin{center}
\scalebox{0.4}{\includegraphics{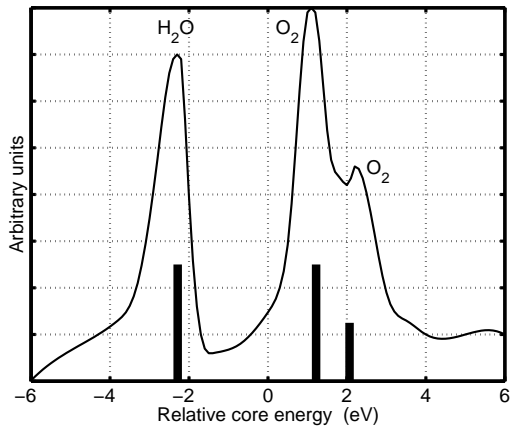}}
\end{center}
\caption{Comparison of wavelet based all-electron calculation of the
location of oxygen $1s$ states in the water and diatomic oxygen
molecules (vertical bars) with experimental ESCA data (curve).
Heights of bars for the O$_2$ molecule are in proportion to the
multiplicity of the spin-state resulting from the photo-emission.  The
height of the bar for H$_2$O is arbitrary. ESCA data are
after~\protect\cite{ESCAbook}.}
\label{fig:core}
\end{figure}

Turning now to the energies associated with these states, we note that
although common approximations to density-functional theory give
relatively poor results for the {\em absolute} positions of deep core
levels, our results for the {\em shifts} in core-level energies show
remarkable agreement with experiment.  Accordingly, in comparing with
experimental values, we maintain the relative placement of all of our
eigenvalues (even between different molecules) by adding a single
constant offset to our $1s$ eigen-energies.  Figure~\ref{fig:core}
compares our multiresolution calculations of the energies of the
oxygen $1s$ levels with the experimentally observed X-ray
photo-emission spectrum of a mixture of water and molecular
oxygen\cite{ESCAbook}.

In molecular oxygen, the experiment shows significant splitting in the
$1s$ levels, with noticeably different cross-sections for the two
levels.  This results from the propagation of paramagnetic bonding
effects in the valence electrons down to the core levels.  The
stronger emission line results from the spin quartet state of the
molecule (the result of photo-emission from the minority spin
channel), which has twice the expected cross-section of the spin
doublet state (the result of photo-emission from the majority spin
channel).  Our calculations, which treat the valence and core physics
on a unified and equal footing, predict precisely this valence-core
effect.  The figure displays our results for the oxygen $1s$ Kohn-Sham
eigenvalues (with a single, constant shift for both molecules) as
vertical bars with heights proportional to the cross-sections expected
for each spin channel.  For the oxygen molecule, we find precisely the
correct splitting and ordering for the two spin states.  Finally, we
note that our calculation of the relative positioning of the oxygen
$1s$ state in the water molecule is also in excellent agreement with
the experiment.  (As the relative strengths of the spectral lines
between oxygen and water depend upon the relative concentrations in
the experiment, the height of the bar in the figure indicating our
calculation of the water molecule is arbitrary.)  With this level of
agreement, such calculations can be a great aid in the interpretation
of spectra of greater complexity.

\section{Conclusions}

Multiresolution all-electron calculations allow the unified treatment
of core and valence electrons where the local-density approximation is
the sole approximation without systematic control.  We have
demonstrated that, with this approach, study of the impact of valence
behaviors on core electrons is straightforward and that meaningful
core polarization effects and core-level shifts may be extracted to
aid in the interpretation of experimental data such as ESCA.

Further, we have shown that, with the use of newly developed
techniques, such calculations may be carried out with an effort
comparable to the corresponding pseudopotential calculations, thereby
establishing for the first time multiresolution analysis as a {\em
viable} approach to all-electron calculations.  By accounting in
detail for the computational demands of our calculations, we have
established a standard against which the performance of future wavelet
calculations should be compared.

\acknowledgments
This work supported by the US DOE ASCI ASAP Level~2 program
(Contract No. B347887).  TDE would like to thank the Research Council
of Norway.  Computational support provided by the Cornell Center for
Materials Research.

\begin{table}
\begin{center}
\begin{tabular}{c|cc|cc}
level &N$_x$ &N$_y$ & $\Delta_x$ & $\Delta_y$ \\  \hline \hline
0     &4    &	4    &-	     &-		\\\hline
1     &4    &	4    &  1	     &	 1	\\\hline
2     &3    &	3    &  1	     &	 1	\\\hline
\end{tabular}
\end{center}
\caption{Numerical representation of restricted multiresolution
analysis in
Figure~\ref{fig:mra2dmonocube}: dimension ($N_x,N_y$) and location
relative to parent ($\Delta_x,\Delta_y$) of each refinement level.}
\label{tbl:nmra2dmonocube}
\end{table}

\begin{table}
\begin{center}
\begin{tabular}{c|c} \hline
$\sigma$ & $Q$ \\
(a.u.) & (a.u.)  \\ \hline \hline
$\frac{1}{8 Z}$ &          $+3.132576693428\, Z$\\
$\frac{1}{4 \sqrt{2} Z}$ & $-2.683558382240\, Z$ \\
$\frac{1}{4 Z}$ &          $+0.550981688812\, Z$
\end{tabular}
\end{center}
\caption{Widths $\sigma$ and norms $Q$ of the three Gaussian charge
distributions $Q \exp(-r^2/(2\sigma^2)) / (\sqrt{2 \pi} \sigma)^3$
superimposed to represent a nucleus of charge $Z$.  All quantities
reported in atomic units.}
\label{tbl:teter}
\end{table}

\begin{table}
\begin{center}
\vspace{0.1cm}
\begin{tabular}{c|ccc|ccc}
level &N$_x$ &N$_y$ &N$_z$ & $\Delta_x$ & $\Delta_y$ & $\Delta_z$ \\  \hline \hline
0     &36    &	36  & 36  &  -	     &	 -	& -	   \\\hline
1     &36    &	36  & 36  &  9	     &	 9	& 9	   \\\hline
2     &36    &	36  & 36  &  9	     &	 9	& 9	   \\\hline
3     &36    &	36  & 36  &  9	     &	 9	& 9	   \\\hline
4     &36    &	36  & 36  &  9	     &	 9	& 9	   \\\hline
5     &36    &	36  & 36  &  9	     &	 9	& 9	   \\\hline
6     &48    &	48  & 48  &  6       &	 6	& 6 \\ \hline
7     &48    &	48  & 48  &  12	     &	 12	& 12 \\ \hline
8     &72    &	72  & 72  &  6       &	 6	& 6 \\ \hline
9     &72    &	72  & 72  &  18	     &	 18	& 18 \\ \hline
\end{tabular}
\end{center}
\caption{Restriction employed for all-electron calculations
of calcium: dimension ($N_x,N_y,N_z$) 
and location relative to parent ($\Delta_x,\Delta_y,\Delta_z$) of 
each refinement level.}
\label{tbl:Cagridstruct}
\end{table}

\begin{table}
\begin{center}
\vspace{0.1cm}
\begin{tabular}{c|ccc|ccc}
level &N$_x$ &N$_y$ &N$_z$ & $\Delta_x$ & $\Delta_y$ & $\Delta_z$ \\  \hline \hline
0     &28    &	24  &24	  &-	     &-		&-	   \\\hline
1     &30    &	24  &24	  &  6	     &	 6	& 6	   \\\hline
2     &34    &	24  &24	  &  6	     &	 6	& 6	   \\\hline
3     &44    &	24  &24	  &  6	     &	 6	& 6	   \\\hline
4     &60    &	24  &24	  &  7	     &	 6	& 6	   \\\hline \hline
5     &24    &	24  &24	  &  6	     &	 6	& 6	   \\\hline
6     &32    &	32  &32	  &  4	     &	 4	& 4	   \\\hline
5     &24    &	24  &24	  &  42	     &	 6	& 6	   \\\hline
6     &32    &	32  &32	  &  4	     &	 4	& 4 \\ \hline
\end{tabular}
\end{center}
\caption{Restrictions employed for all-electron calculation of oxygen
molecule: dimension ($N_x,N_y,N_z$) and origin relative to its
parent ($\Delta_x,\Delta_y,\Delta_z$) of each refinement level.}
\label{tbl:gridsO2}
\end{table}

\begin{table}
\begin{center}
\vspace{0.1cm}
\begin{tabular}{c|ccc|ccc}
level &N$_x$ &N$_y$ &N$_z$ & $\Delta_x$ & $\Delta_y$ & $\Delta_z$ \\  \hline \hline
0     &48    &	48  &48	  &  -	     & - &	-   \\\hline
1     &48    &	48  &48	  &  12	     &	 12	& 12	   \\\hline
2     &48    &	48  &48	  &  12	     &	 12	& 12	   \\\hline
3     &72    &	72  &48	  &  6	     &	 6	& 12	   \\\hline
4     &72    &	72  &48	  &  18	     &	 18	& 12	   \\\hline
5     &72    &	72  &48	  &  18	     &	 18	& 12	   \\\hline
6     &72    &	72  &48	  &  18	     &	 18	& 12	   \\\hline
\end{tabular}
\end{center}
\caption{Restrictions employed for all-electron calculation of 
water molecule: dimension ($N_x,N_y,N_z$) and origin relative to its
parent ($\Delta_x,\Delta_y,\Delta_z$) of each refinement level.}
\label{tbl:gridsH2O}
\end{table}

\begin{table}
\begin{center}
\begin{tabular}{l|c|c}
Quantity & MRA & $V_{ps}$\\ \hline\hline
$N_B$ & 200,000 & 160,000 \\ \hline
$N_{F/B}$ & 250 & 1600 \\ \hline
$N_{F/T}$ & 200$\cdot 10^6$ & 180$\cdot 10^6$ \\ \hline
$N_A$ & 32,000 & 8,500 \\ \hline
$T$  & 8.2~hrs & 3.4~hrs \\ \hline
\end{tabular}
\end{center}
\caption{Analysis and comparison of computational demands of
multiresolution analysis all-electron calculation (MRA) and plane-wave
pseudopotential calculation ($V_{ps}$) of oxygen molecule within LSDA
for basis sets reaching a similar level of convergence.  }
\label{tbl:performance}
\end{table}

\end{document}